\documentclass{ws-ijmpe}
\usepackage[super,compress]{cite}
\usepackage{amsmath,amssymb}
\usepackage{color}

\begin{document}
        
\markboth{A. M. Shirokov, A. I. Mazur, I. A. Mazur, V. A. Kulikov}{Tetraneutron resonances and its isospin analogues}

\catchline{}{}{}{}{}
        
\title{Tetraneutron resonance and its isospin analogues}

\author{A. M. Shirokov}
\address{Skobeltsyn Institute of Nuclear Physics, Lomonosov Moscow State University, Moscow 119991, Russia\\
shirokov@nucl-th.sinp.msu.ru}

\author{A. I. Mazur}
\address{Laboratory for Modeling of Quantum Processes, Pacific National University, Khabarovsk 680035, Russia\\
mazur@khb.ru}

\author{I. A. Mazur}
\address{Center for Exotic Nuclear Studies, Institute for Basic Science, Daejeon 
34126, Republic of Korea\\
mazuri@ibs.re.kr}

\author{V. A. Kulikov}
\address{Skobeltsyn Institute of Nuclear Physics, Lomonosov Moscow State University, Moscow 119991, Russia\\
kulikov@nucl-th.sinp.msu.ru}

\maketitle

\begin{history}
\received{}
\revised{}
\end{history}

\begin{abstract}
We discuss  isospin analogues of the tetraneutron resonance and their possible
manifestations in nuclear structure and reactions.
\end{abstract}

\keywords{Tetraneutron; isospin analogues}
\ccode{PACS numbers: 21.45.+v, 21.10.Hw, 23.90.+w, 27.10.+h}

\mbox{}

Studies of light multineutron systems and, in particular, system of four neutrons (tetraneutron, $^{4}n$)
have been performed within more than 60 years. 
Experimental searches for a bound or resonant tetraneutron state
were started in early 1960s in pion double-charge exchange reactions on $^{4}$He, in
nuclear fission
and later in multinucleon transfer reactions. The historical overview of respective experimental 
investigations and related theoretical efforts can be found in 
Refs.~\refcite{Roman,Miguel-review}.
For a long time no definite signals of tetraneutron were found. In 2002, the GANIL 
experiment~\cite{Miguel-2002} reported the bound tetraneutron detected in the 
${\rm^{14}Be\to{^{10}Be}}+{^4n}$ breakup on a carbon target. This result was not confirmed in
later experiments, however, it triggered a number of theoretical studies of tetraneutron
(see reviews~\refcite{Roman,Miguel-review}). In particular, these studies confirmed 
 using modern internucleon interactions and theoretical methods the 
earlier conclusion\cite{BGGZ} based on the nuclear instability of $^{5}$H that four neutrons cannot be 
bound.
A number of
theoretical efforts performed in various techniques allowing for the continuum
(Faddeev--Yakubovsky calculations, complex scaling method, analytic continuation on
the coupling constant, etc.) with various modern $NN$ interactions like CD-Bonn,
Argonne AV18 or chiral
effective field theory ($\chi$EFT) interactions with and without  additional
three-nucleon ($3N$) forces
failed to obtain a low-energy tetraneutron resonance narrow enough to be
detected in experiment. We note here that conventional $3N$ forces fitted to three-nucleon
bound states are characterized by the isospin $T=1/2$ while 
the tetraneutron requires the $3N$ force 
with isospin $T=3/2$ which parameters are unknown.

The situation changed in 2016 when the tetraneutron resonance at the energy
of $E_r=0.83\pm 0.65({\rm stat})\pm1.25({\rm syst})$~MeV and the width $\Gamma<2.6$~MeV
was detected in the ${\rm{^4He}({^8He}},2\alpha){^4n}$  reaction~\cite{Kisamori}. The large 
 uncertainties of the resonance energy and width are due to
a marginal statistics of 4~events. In the same 2016, the tetraneutron resonance was first
predicted in Ref.~\refcite{PRL2016} at the energy of $E_r=0.844$~MeV with the 
width $\Gamma=1.378$~MeV in calculations with the  JISP16 $NN$
interaction~\cite{JISP16} using
the {\it ab initio} no-core shell model (NCSM)~\cite{NCSM} extended to the four-body
continuum within the minimal democratic decay approximation by means of the
SS-HORSE method~\cite{SS-HORSE,SS-HORSE-EChAYa}. Later similar predictions were
obtained in the same approach~\cite{AIP-2018} with the $NN$ interactions
Daejeon16~\cite{Daejeon16}
and  $\chi$EFT Idaho N3LO~\cite{Idaho} softened by the SRG transformation without
$3N$ forces. 

In our calculations~\cite{PRL2016,AIP-2018}, 
the tetraneutron continuum is described within the democratic decay 
approximation,  and thus the system disintegrating 
 into 4 neutrons should penetrate through a high centrifugal barrier arising in 
 the  hyperspherical harmonics method which is natural for modeling the democratic decay.
One can argue that this high hyperspherical 
 centrifugal barrier can  decrease the resonance width or just create artificially the 
 resonance while the system may dynamically create two-body channels of the type
 $^{2}n+{^{2}n}$ or $n+{^{3}n}$ to facilitate the decay. To check this idea, we  performed the
NCSM--SS-HORSE  calculations with the Daejeon16 interaction of the tetraneutron decaying
in the $^{2}n+{^{2}n}$ or the $n+{^{3}n}$ 
channels. In the latter case we used the results of our calculations of the trineutron 
resonance~\cite{3n}. The preliminary results of these calculations 
confirmed the consistency of the democratic
decay approximation: the tetraneutron resonance energy obtained with the two-body channels was
similar to that of the democratic
decay approximation while the resonance width obtained with two-body channels was not
smaller than the democratic
decay width.

It should be noted that, as various other theoretical studies, our calculations of
Ref.~\refcite{AIP-2018} did not predict the tetraneutron resonance if the initial non-softened
 Idaho N3LO $NN$ interaction is used.
However, according to Ref.~\refcite{AIP-2018}, Idaho N3LO $NN$ interaction supports
a low-lying virtual state  in the four-neutron system which may generate some
resonance-like signals in reactions producing four neutrons. 

The same Idaho N3LO $NN$ 
interaction but renormalized by  the $V_{{\rm low}\mbox{-}k}$
method was utilized in the no-core Gamow shell model (NCGSM) calculations of
Ref.~\refcite{Michel} which resulted in the prediction for the tetraneutron resonance at the energy
of~${E_{r}=2.64}$~MeV with the width~$\Gamma=2.38$~MeV. 

The final confirmation
of the existence of the tetraneutron resonance was obtained in the 2022 experimental
study of the ${\rm^{8}He}(p,p{\rm^{4}He}){^{4}n}$ reaction
with a reasonable statistics~\cite{Duer} where the resonance energy and width were estimated
respectively as ${E_{r}= 2.37\pm 0.38({\rm stat}) \pm 0.44({\rm syst})}$~MeV and 
${\Gamma= 1.75 \pm0.22({\rm stat})\pm 0.30({\rm syst})}$~MeV.

However, some physicists are still  in a doubt about the existence of the tetraneutron
resonance~\cite{Sobotka,Lazauskas,Dubna}. In particular, Sobotka and Piarulli~\cite{Sobotka}
suppose that another interpretation of the results of the experiment~\cite{Duer} is possible
related to the mechanism of the ${\rm^{8}He}(p,p{\rm^{4}He}){^{4}n}$ reaction without 
discussing the details of this mechanism. Lazauskas
{\it et al.}~\cite{Lazauskas} studied the ${\rm^{8}He}(p,p{\rm^{4}He}){^{4}n}$ reaction
using the method 
of ``sudden removal'' of the ${\rm^{4}He}$ core from the ${\rm^{8}He}$
nucleus suggested in Ref.~\refcite{Grigorenko}. 
They performed Faddeev--Yakubovsky calculations of four neutrons surrounding
the ${\rm^{4}He}$ core of an infinite mass. As a result, the center-of-mass treatment
in the ${\rm^{8}He}$ wave function is incorrect and was 
criticized in Ref.~\refcite{Dubna}. It is interesting that the authors of Ref.~\refcite{Dubna}
performed an experimental study of the reactions
${\rm{^2He}({^8He},{^6Li}}){^4n}$ and  ${\rm{^2He}({^8He},{^3He}){^7H}\to{^3H}} + {^4n}$  and
got the  peaks consistent with the tetraneutron resonance observed in the
experiment of Ref.~\refcite{Duer}. However they
prefer to interpret their results as a complicated reaction mechanism supported by their 
calculations of $^8$He in a five-body
$\alpha+{^4n}$ and of $^4n$ in a four-body hyperspherical harmonics models avoiding the 
formation of the tetraneutron resonance.
The idea of this interpretation is explained in Ref.~\refcite{Dubna}:
``{\it The existence of low-energy tetraneutron resonance would
mean a radical revision of everything we know about neutron-rich 
nuclei and neutron matter. Our vision of the problem is
that a solution can be found, which is much less radical being
related to the $^8$He structure and reaction mechanisms.}''

We suppose that the skepticism  regarding the existence of the tetraneutron resonance
is supported by the failure to obtain it in
numerous theoretical studies accounting for the effects of the continuum with modern 
internucleon forces (see reviews~\refcite{Roman,Miguel-review}). We note here again that the
tetraneutron calculations require the $3N$ force 
with isospin $T=3/2$ which is unknown. Summarizing the theoretical efforts, the tetraneutron
resonance was predicted only with $\chi$EFT $NN$ interactions softened by
SRG~\cite{AIP-2018} or $V_{{\rm low}\mbox{-}k}$~\cite{Michel} methods or in calculations
with JISP16~\cite{PRL2016} and Daejeon16~\cite{AIP-2018} $NN$ interactions. The softening
of the $NN$ interactions generally induces the $3N$ force which may cancel the ``generic''
 $T=3/2$ $3N$ interaction but there is no any study supporting this cancellation. More honest
are the results obtained with the JISP16  and Daejeon16  interactions. These interactions are
realistic, i.\,e., they describe the deuteron properties and $NN$ scattering data, and are fitted 
to the description of $s$- and $p$-shell nuclei by off-shell modifications that mimic the
effect of $3N$ forces~\cite{JISP16,Daejeon16}. The  JISP16  and Daejeon16 
were extensively utilized in many-body applications of various groups since they make it 
possible to avoid using the $3N$ interactions which essentially simplifies the calculations. 
These interactions, especially Daejeon16, provide an accurate description of $p$-shell
nuclei including exotic neutron-excess isotopes~\cite{Daejeon16,NN-NNN,rotational}, their  binding
energies and spectra including collective phenomena like rotational bands~\cite{rotational}, 
rms radii~\cite{rodkin} and electromagnetic transitions~\cite{Li}, etc. JISP16 was used to 
predict~\cite{14F} properties of the $^{14}$F nucleus; the accuracy of this prediction was
later confirmed by an experiment~\cite{Gold} where this isotope was first observed.
However, the  failure of modern $\chi$EFT $NN$ forces to predict the tetraneutron resonance is
not in line with the modern trend to  use them for describing nuclear properties. 

It is interesting that the same JISP16, Daejeon16, SRG-softened and 
\mbox{$V_{{\rm low}\mbox{-}k}$-softened}
Idaho N3LO $NN$ interactions support the trineutron resonance below the
tetraneutron one~\cite{3n,Michel} while the calculations with non-softened $\chi$EFT Idaho N3LO and 
LENPIC N4LO semilocal coordinate
space $NN$ interactions~\cite{LENPIC} did not predict the trineutron resonance~\cite{3n} as well
as calculations allowing for the continuum spectrum effects
with other modern nuclear forces~\cite{Roman,Miguel-review}.

In order either to confirm or reject the existence of the tetraneutron resonance, we
suggest to study its isospin analogues. The usual nuclear physics wisdom suggests that 
the tetraneutron resonance is a member of the  isospin multiplet which should also
include the excited resonant $0^+,T=2$ states in $^{4}$H, $^{4}$He, $^{4}$Li nuclei and a 
tetraproton resonance 
 which is the unbound $^{4}$Be isotope. The experimental search for these resonances is
of interest from various points of view. 

Assuming the  charge independence of nuclear forces, we expect the $0^+,T=2$ resonance in 
$^{4}$H at the same energy and with the same width as the tetraneutron resonance, i.\,e.,
accepting the values from the experiment~\cite{Duer}, at the energy ${E_{r}\approx 2.4}$~MeV
above the threshold of the $^{4}$H disintegration into 4 nucleons. The $^{4}$H nucleus is
unbound, its binding energy is ${5.6\pm0.1}$~MeV~\cite{cdfe}; hence the $0^+,T=2$ resonance is 
expected at the excitation energy of $E_{x}\approx 8$~MeV with the 
width $\Gamma\approx 1.8$~MeV. Thus the  $0^+,T=2$ resonance in 
$^{4}$H is expected above its already known excited states~\cite{cdfe}: 
$1^{-},T=1$ (${E_{x}=0.31}$~MeV,
${\Gamma=6.73}$~MeV), ${0^-,T=1}$ ($E_{x}=2.08$~MeV, ${\Gamma=8.92}$~MeV), and
$1^{-},T=1$ (${E_{x}=2.83}$~MeV, ${\Gamma=12.99}$~MeV). The desired $0^+,T=2$ state
in $^{4}$H can be populated in various nuclear reactions, for example, in the
$\alpha$-transfer reaction ${\rm{^{8}Li}({^{14}C},{^{18}O}){^{4}H}}^{*}(0^+,T = 2)$.

The mirror $^{4}$Li nucleus is also unbound, its binding energy is $4.6\pm0.2$~MeV~\cite{cdfe}.
The known excited states in $^{4}$Li have very close excitation energies to those in  $^{4}$H and
slightly larger widths~\cite{cdfe}: $1^{-},T=1$ (${E_{x}=0.32}$~MeV,
${\Gamma=7.35}$~MeV), ${0^-,T=1}$ ($E_{x}=2.08$~MeV, ${\Gamma=9.35}$~MeV), and
$1^{-},T=1$ (${E_{x}=2.85}$~MeV, ${\Gamma=13.51}$~MeV). Therefore we can
expect the desired $0^+,T=2$ resonance in 
$^{4}$Li at the same excitation energy of $E_{x}\approx 8$~MeV and with a slightly larger
width of $\Gamma\approx 2$~MeV. This resonance can be populated in various
nuclear reactions. Note that $p+\rm{^{3}He}$ collision which is a natural and convenient
 reaction for studying experimentally the $^{4}$Li states,  is isospin forbidden
for creating the  $T=2$ state in  $^{4}$Li. If, however, the $0^+,T=2$ resonance will be detected
in the $p+\rm{^{3}He}$ reaction with a small cross section, it will be an interesting direct
observation of the isospin violation.

The tetraproton ($^{4}$Be nucleus) would be complicated to create in nuclear reactions.
It is supposed to be the $0^+,T=2$ resonance  at the energy of around 3.5~MeV and with the width
of about 2.5~MeV.

The $^{4}$He nucleus is well studied both experimentally and theoretically, however, the
$T=2$ states were never observed in it. Due to the large binding energy of 
28.296~MeV~\cite{cdfe}, the $0^{+},T=2$ state in   $^{4}$He is expected at the excitation
energy of approximately 31~MeV. At this energy there is a large density of wide overlapping
states in  $^{4}$He. Therefore we suppose 
that the direct experimental observation of the wide $0^{+},T=2$ 
resonance in $^{4}$He would be very difficult.

However the $0^{+},T=2$ state in   $^{4}$He have been probably already observed indirectly. Suppose that
an $\alpha$ particle in the ground state is colliding with another $\alpha$ particle excited to
the $0^{+},T=2$ state of our interest, i.\,e., we have the ${\rm{^{4}He}+{^{4}He}}^{*}(0^{+},T=2)$
system. This is an excited state of ${\rm{^{8}Be}}^{*}(0^{+},T=2)$. This very narrow state
with the width $\Gamma=5.5$~keV have been observed in ${\rm{^{8}Be}}$ at the excitation energy
of 27.494~MeV~\cite{cdfe}. It looks like that the presence of another $\alpha$ particle
lowers down the ${\rm{^{4}He}}^{*}(0^{+},T=2)$ excited state by 
approximately 3.5~MeV
as compared with
our above analysis based on the central value for the tetraneutron energy
from the experimental study of Ref.~\refcite{Duer}. However, noting 
the statistical and systematic 
experimental uncertainties,  the accuracy of this estimate  is about 
1~MeV. 

The ${\rm{^{8}Be}}^{*}(0^{+},T=2)$ state is, of course, the isospin analogue of the $^{8}$He
ground state. Supposing that ${\rm{^{8}Be}}^{*}(0^{+},T=2)$ has a structure of
${\rm{^{4}He}+{^{4}He}}^{*}(0^{+},T=2)$, we should expect that  the $^{8}$He 
wave function has a
component ${\rm{^{4}He}}+{^{4}n}$. We note that the $^{6}$He wave function has a strong
contribution of the component with the dineutron ($^{2}n$), i.\,e.
${\rm{^{4}He}}+{^{2}n}$  (see, e.\,g., Ref.~\refcite{6He}). The dineutron is unbound (it
has a low-lying virtual state) but produces nearly a 1~MeV of additional binding
by forming the   $^{6}$He ground state in  the
 presence of $\alpha$ particle. The tetraneutron is also unbound but produces approximately
3.1~MeV of additional binding in the presence of $\alpha$ particle when forming the 
$^{8}$He nucleus. 
However, the ${{\rm{^{4}He}}+{^{4}n}}$ is 
only one of the components in the $^{8}$He ground state
wave function and 
a fraction of the 3.1~MeV binding in $^{8}$He is expected to be produced by other
components of the wave function.

In the case of the ${\rm{^{8}Be}}^{*}(0^{+},T=2)$ state, the excited 
${\rm{^{4}He}}^{*}(0^{+},T=2)$ state  in the
presence of $\alpha$ particle produces approximately 0.8~MeV  binding relative to the
${\rm{^{4}He}+p+p+n+n}$ decay threshold. It is smaller than tetraneutron 
in the case of $^{8}$He but note the Coulomb
repulsion. We can also expect that due to the diluted two-$\alpha$ structure of $^{8}$Be,
one of the $\alpha$ particles takes all $T=2$ excitation 
and the ${\rm{^{4}He}+{^{4}He}}^{*}(0^{+},T=2)$ component in ${\rm{^{8}Be}}^{*}(0^{+},T=2)$ is
enhanced as compared to the  ${\rm{^{4}He}}+{^{4}n}$ component in $^{8}$He thus
reducing the contribution  to the binding.of other components of the wave function.

If the 
 ${\rm{^{4}He}+{^{4}He}}^{*}(0^{+},T=2)$ interpretation of the ${\rm{^{8}Be}}^{*}(0^{+},T=2)$
state is correct, we have an interesting situation: the ${\rm{^{4}He}}^{*}(0^{+},T=2)$ can
experience only a complete four-nucleon decay due to its isospin but the  
${\rm{^{8}Be}}^{*}(0^{+},T=2)$ resonance energy of 27.494~MeV is not enough for the 
disintegration in the channel ${\rm{^{4}He}+p+p+n+n}$ and this state
decays through some other channels. For example, ${\rm{^{8}Be}}^{*}(0^{+},T=2)$ can
emit one or two protons. In the first case, due to the isospin conservation, we can expect the
decay  ${{\rm{^{8}Be}}^{*}(0^{+},T=2)\to {\rm{^{7}Li}}^{*}(3/2^{-},T=3/2)+p}$ which requires 
$p$ wave in the ${p{-}{\rm{^{7}Li}}^{*}(3/2^{-},T=3/2)}$ relative motion. The $^{7}$Li
binding energy is 39.245~MeV, the ${\rm{^{7}Li}}^{*}(3/2^{-},T=3/2)$ state has an excitation energy
of approximately 11.2~MeV~\cite{cdfe}; thus the proton is emitted by  
${\rm{^{8}Be}}^{*}(0^{+},T=2)$ at the energy of approximately 650~keV and this decay is 
suppressed by the centrifugal and Coulomb barriers. The two-proton decay
${{\rm{^{8}Be}}^{*}(0^{+},T=2)\to {\rm{^{6}He}}+p+p}$ happens at the energy of approximately
1.9~MeV and is energetically more preferable. However, an energy per proton is only slightly
larger than in the previous case and, due to the naive shell model 
expectations, each proton is emitted from the $p$ state and also experiences 
the centrifugal and Coulomb barriers. Note also an essential rearrangement of the
 ${\rm{^{4}He}+{^{4}He}}^{*}(0^{+},T=2)$ structure in the cases of the one- or two-proton
decays of ${\rm{^{8}Be}}^{*}(0^{+},T=2)$ that results in their additional suppression.
One more option is a decay through some channels
 violating the isospin conservation due to a small admixture of the $T\ne 2$ components in the
${\rm{^{8}Be}}^{*}(0^{+},T=2)$ wave function as was discussed in Ref.~\refcite{BK}. These
considerations explain
the narrowness of the ${\rm{^{8}Be}}^{*}(0^{+},T=2)$ resonance.


The suggested ${\rm{^{4}He}+{^{4}He}}^{*}(0^{+},T=2)$  structure of the 
${\rm{^{8}Be}}^{*}(0^{+},T=2)$ state is {strongly supported} by 
the excited $0^{+},T=2$ state in  $^{12}$C which can be interpreted
as the ${{\rm{^{8}Be}+{^{4}He}}^{*}(0^{+},T=2)}$ system. The excitation energy of the
${{\rm{^{12}C}}^{*}(0^{+},T=2)}$ resonance is $E_{x}=27.595$~MeV and its width is
${\Gamma<30}$~keV~\cite{cdfe}. Note, this excitation energy is only 100~keV
larger than that of the ${\rm{^{8}Be}}^{*}(0^{+},T=2)$ state which is consistent with the
fact that the $^{8}$Be ground state is underbound by 92~keV with respect to the
$\alpha+\alpha$ threshold. This surprising nearly exact coincidence of the excitation energies of 
${\rm{^{8}Be}}^{*}(0^{+},T=2)$ and ${{\rm{^{12}C}}^{*}(0^{+},T=2)}$ states together with the
smallness of their widths confirms 
our interpretation of their cluster structure including the ${\rm{^{4}He}}^{*}(0^{+},T=2)$ state.

The $T=2$ excited state in $^{12}$C at the energy of 29.63~MeV and the width
$\Gamma<200$~keV has a  tentative spin-parity assignment $2^{+}$~\cite{cdfe}.
This ${{\rm{^{12}C}}^{*}((2^{+}),T=2)}$ state can be probably interpreted as a
${{{\rm{^{8}Be}}^{*}(2^{+},T=0)+{\rm^{4}He}}^{*}(0^{+},T=2)}$ system where the
${{{\rm{^{8}Be}}^{*}(2^{+},T=0)}}$ state has an excitation energy of 3.03~MeV~\cite{cdfe}.

The $0^{+},T=2$ state in  $^{16}$O has an excitation energy of $E_{x}=22.721$~MeV
and the width ${\Gamma=12.5}$~keV~\cite{cdfe}. Having in mind that the $^{12}$C nucleus
is bound by 7.275~MeV with respect to the $3\alpha$ threshold, an interpretation of the
${{\rm{^{16}O}}^{*}(0^{+},T=2)}$ state as the ${{\rm{^{12}C}+{^{4}He}}^{*}(0^{+},T=2)}$ system
suggests a strange conclusion
that the ${{\rm^{4}He}^{*}(0^{+},T=2)}$ is by a couple of MeV
less bound in the presence of $^{12}$C
than in the presence of $^{4}$He or $^{8}$Be. An interpretation of the
${\rm{^{16}O}}^{*}(2^{+},T=2)$ state with the excitation energy $E_{x}=24.522$~MeV
and width ${\Gamma<50}$~keV as the ${{\rm{^{12}C}^{*}(2^{+},T=0)+{^{4}He}}^{*}(0^{+},T=2)}$ system
where the  excitation energy of ${{\rm{^{12}C}^{*}(2^{+},T=0)}}$ is 
$E_{x}=4.44$~MeV~\cite{cdfe}, is 
consistent with the estimation of the
${\rm^{4}He}^{*}(0^{+},T=2)$ energy in the ${{\rm{^{8}Be}}^{*}(0^{+},T=2)}$ and
${{\rm{^{12}C}}^{*}(0^{+},T=2)}$ resonant states.

\mbox{}

{\it Conclusion.} We discussed the tetraneutron resonance, its theoretical studies and the
experimental discovery as well as some concerns regarding the interpretation of the
results of the respective experiment. We suppose it will be interesting to search for the
tetraneutron isospin analogues in the excited $T=2$ states of  $^{4}$H, $^{4}$He, $^{4}$Li and
in the ground state of $^{4}$Be (tetraproton). We suppose that unbound
excited $0^{+},T=2$ states in $^{4}$H and $^{4}$Li have a better prospect for the direct
experimental investigation. The excited $0^{+},T=2$ state in $^{4}$He is expected at the excitation
energies where there is a high density of wide overlapping states and thus it will be hard to
separate this state. However this ${\rm^{4}He}^{*}(0^{+},T=2)$ state seems to be manifested
in highly excited narrow states of $^{8}$Be, $^{12}$C and $^{16}$O nuclei.

\mbox{}

This research is supported by the Russian Science Foundation  
Grant No~24-22-00276.


\begin{thebibliography}{99}

\bibitem{Roman} R. Kezerashvili, {\it Fission and Properties of Neutron-Rich Nuclei}
(World Scientific, Singapore, 2017), p.~403; arXiv:1608.00169 [nucl-th] (2016).

\bibitem{Miguel-review} F. M. Marqu\'es and J. Carbonell, {\it Eur. Phys. J. A} {\bf 57}, 105 (2021).

\bibitem{Miguel-2002} F. M. Marqu\'es {\it et al.}, {\it Phys. Rev. C} 65, 044006 (2002).

\bibitem{BGGZ}A. I. Baz', V. I. Goldansky, V. Z. Goldberg, and Ya. B. Zel'dovich,
{\it Light and Intermediate Nuclei Near the Drip Lines} (Nauka, Moscow, 1972) ({\it in Russian}).


\bibitem{Kisamori}K. Kisamori {\it et al.}, {\it Phys. Rev. Lett.} {\bf 116}, 052501 (2016).

\bibitem{PRL2016}A. M. Shirokov, G. Papadimitriou, A. I. Mazur, I.~A.~Mazur, 
R.~Roth, and J.~P.~Vary, {\it Phys. Rev. Lett.} {\bf 117}, 182502 (2016).


\bibitem{JISP16} A. M. Shirokov, J. P. Vary, A. I. Mazur, and T. A. Weber,
{\it Phys. Lett. B} {\bf644}, 33 (2007); a Fortran code generating the
JISP16 matrix elements is available at 

\bibitem{NCSM}B. R. Barrett, P. Navr\'atil, and J. P. Vary, {\it Prog. Part. Nucl. Phys.}
{\bf69}, 131 (2013).

\bibitem{SS-HORSE}A. M. Shirokov, A. I. Mazur, I. A. Mazur, and J. P. Vary,
{\it Phys. Rev. C} {\bf94}, 064320 (2016).

\bibitem{SS-HORSE-EChAYa}I. A. Mazur, A. M. Shirokov, A. I. Mazur, and J. P. Vary,
{\it Phys. Part. Nuclei} {\bf48}, 84 (2017).

\bibitem{AIP-2018}A. M. Shirokov, Y. Kim, A. I. Mazur, I. A. Mazur, I.~J.~Shin,
and J. P. Vary, {\it AIP Conf. Proc.} {\bf2038}, 020038 (2018).

\bibitem{Daejeon16}A. M. Shirokov, I. J. Shin, Y. Kim, M. Sosonkina, P.~Maris,
and J. P. Vary, {\it Phys. Lett. B} {\bf761}, 87 (2016); a Fortran code
generating the Daejeon16 matrix elements is available at

\bibitem{Idaho}R. Machleidt and D. R. Entem, {\it Phys. Rev. C} {\bf68}, 041001 (2003).


\bibitem{3n} I. A. Mazur, M. K. Efimenko, A. I. Mazur, I. J. Shin, V.~A.~Kulikov, A. M. Shirokov,
and J. P. Vary, {\it Phys. Rev. C} {\bf110}, 014004 (2024).

\bibitem{Michel}J. G. Li, N. Michel, B. S. Hu, W. Zuo, and F. R. Xu, {\it Phys. Rev.
C} {\bf100}, 054313 (2019).

\bibitem{Duer}M. Duer {\it et al.}, {\it Nature} {\bf606}, 678 (2022).

\bibitem{Sobotka}L. G. Sobotka and M. Piarulli, {\it Nature} {\bf606}, 656 (2022).
\bibitem{Lazauskas} R. Lazauskas, E. Hiyama, and J. Carbonell, {\it Phys. Rev. Lett.} {\bf130}, 102501 (2023).
\bibitem{Dubna} I. A. Muzalevskii {\it et al.}, 
arXiv:2312.17354 [nucl-ex] (2023).

\bibitem{Grigorenko} L. V. Grigorenko, N. K. Timofeyuk, and M. V. Zhukov, {\it Eur.
Phys. J. A} {\bf19}, 187 (2004).
%

\bibitem{NN-NNN} A. M. Shirokov, V. A. Kulikov, P. Maris, and J. P. Vary, in
{\it $NN$ and $3N$ Interactions}, edited by L. D. Blokhintsev and I. I.
Strakovsky (Nova Science, Hauppauge, NY, 2017), Chap. 8,
p. 231; 
\verb+https://novapublishers.com/wp-content/uploads/2019/+
\verb+05/978-1-63321-053-0_ch8.pdf+.

\bibitem{rotational}P. Maris, M. A. Caprio, and J. P. Vary, {\it Phys. Rev.
C} {\bf91}, 014310 (2015).

\bibitem{rodkin} D. M. Rodkin  and Yu. M. Tchuvil'sky, {\it Phys. Rev.
C} {\bf106}, 034305 (2022).

\bibitem{Li}H. Li,  D. Fang, H. J. Ong, A. M. Shirokov, J. P. Vary, P. Yin, and X. Zhao, 
arXiv:2401.05776  [nucl-ex] (2024), {\it submitted to Phys. Rev. C}.

\bibitem{14F}P. Maris, A. M. Shirokov, and J. P. Vary, {\it Phys. Rev. C} {\bf81} 021301 (2010). 

\bibitem{Gold}V. Z. Goldberg {\it et al.}, 
{\it Phys. Lett. B} {\bf 692},  307 (2010).

\bibitem{LENPIC}E. Epelbaum, H. Krebs, and U.-G. Mei\ss ner, {\it Phys. Rev. Lett.}
{\bf115}, 122301 (2015).

\bibitem{cdfe}

\bibitem{6He}B. V. Danilin, V. D. Efros, J. S. Vaagen, M. V. Zhukov, and  I. J. Thompson,
{\it Phys. Rep.} {\bf 264}, 27 (1996).

\bibitem{BK} F. C. Barker and N. Kumar, {\it Phys.Lett. B} {\bf30}, 103 (1969).

\end{thebibliography}
\end{document}